\documentclass[sigconf,authorversion,nonacm]{acmart}

\acmConference[ICSE 2024]{46th International Conference on Software Engineering}{April 2024}{Lisbon, Portugal}
\acmBooktitle{Woodstock '18: ACM Symposium on Neural Gaze Detection,
 June 03--05, 2018, Woodstock, NY} 
\acmPrice{15.00}
\acmISBN{978-1-4503-XXXX-X/18/06}

\usepackage{soul}
\usepackage{listings}
\begin{document}

\title{Commit Messages in the Age of Large Language Models}

\author{Cristina V. Lopes}
\email{lopes@uci.edu}
\affiliation{%
  \institution{University of California, Irvine}
  \city{Irvine}
  \state{CA}
  \country{USA}
  \postcode{92697}
}

\author{Vanessa I. Klotzman}
\email{vklotzma@uci.edu}
\affiliation{%
  \institution{University of California, Irvine}
  \city{Irvine}
  \state{CA}
  \country{USA}
  \postcode{92697}
}

\author{Iris Ma}
\email{huaiyaom@uci.edu}
\affiliation{%
  \institution{University of California, Irvine}
  \city{Irvine}
  \state{CA}
  \country{USA}
  \postcode{92697}
}

\author{Iftekar Ahmed}
\email{iftekha@uci.edu}
% \authornotemark[]
\affiliation{%
  \institution{University of California, Irvine}
  \city{Irvine}
  \state{CA}
  \country{USA}
  \postcode{92697}
}

\renewcommand{\shortauthors}{Lopes and Klotzman, et al.}

\begin{abstract}
Commit messages are explanations of changes made to a codebase that are stored in version control systems. They help developers understand the codebase as it evolves. However, writing commit messages can be tedious and inconsistent among developers. To address this issue, researchers have tried using different methods to automatically generate commit messages, including rule-based, retrieval-based, and learning-based approaches. Advances in large language models offer new possibilities for generating commit messages. In this study, we evaluate the performance of OpenAI's ChatGPT for generating commit messages based on code changes. We compare the results obtained with ChatGPT to previous automatic commit message generation methods that have been trained specifically on commit data. Our goal is to assess the extent to which large pre-trained language models can generate commit messages that are both quantitatively and qualitatively acceptable. We found that ChatGPT was able to outperform previous Automatic Commit Message Generation (ACMG) methods by orders of magnitude, and that, generally, the messages it generates are both accurate and of high-quality. We also provide insights, and a categorization, for the cases where it fails.
\end{abstract}





\received{20 February 2007}
\received[revised]{12 March 2009}
\received[accepted]{5 June 2009}

\newcommand{\acmg}{automatic commit message generation}
\maketitle

\section{Introduction}
Code is constantly being written and modified on a daily basis. To keep track of these changes, developers use version control systems. In these systems, every alteration is accompanied by a commit message, which is a brief explanation of what the change is. The version control system saves this commit message, along with the ``diffs'' (differences) between the current version and the previous version of the code. Commit messages and the changes they represent are crucial for understanding and maintaining software. They provide insight into what changes were made and why, which can help developers better understand the design and architectural decisions that were made. Clear and informative commit messages can save software developers time by providing a clear understanding of the reasoning behind changes. They also aid in tracking the evolution of the codebase over time.

While there is no set rule for what should be included in a commit message, a good commit message should provide context for the changes made, such as the reasons for the change, any related issues, or potential consequences. However, crafting a meaningful commit message takes effort, discipline, and time, which developers may not always have. This can lead to poor quality commit messages, such as messages that are missing, too brief, or simply repeat information found in the diffs. A study by Dyer et al.~\cite{10.5555/2486788.2486844}  found that 14\% of commit messages in over 23,000 open-source Java projects on SourceForge were empty and 75\% had fewer than 16 words. Effective communication among developers is crucial for successful collaborative software development. Commit messages serve as a record of how, by whom, and why the source code of a project has changed. The quality of these messages greatly affects communication among developers, with poor quality messages hindering it. 

To improve commit messages, researchers have been exploring the idea of automatically generating them by means of using rule-, retrieval-, and, more recently, learning-based approaches~\cite{10.1145/1858996.1859005,CortesCoy2014OnAG,7203049,7551998,huang2020learning,9000000,8115626,loyola2017neural,Xu2019CommitMG,10.1145/3328833.3328873,Liu2019GeneratingCM,nie2021coregen,jung2021commitbert,Pavel2021ASN,Rastkar2013WhyDT,liu2020atom,9425953,10.1145/3464689,exemplar,10.1109/ESEM.2017.56,9793882,https://doi.org/10.48550/arxiv.2203.02700}. Rule-based approaches do not fully account for all situations, and cannot clearly convey the reasoning or intent behind code changes. Retrieval-based approaches tend to produce inaccurate results if there are no similar code changes in the dataset. Learning-based approaches seem to be the most promising ones. Over the past 5 years, there has been a steady stream of work describing supervised machine learning models for Automatic Commit Message Generation (ACMG). All that work has been focused on commit data in isolation: the models were trained on datasets of diffs and the corresponding commit messages. While the results are better than previous approaches, the quality and correctness of the resulting messages still leaves much room for improvement.

The recent emergence of Large Language Models (LLM) trained on extremely large datasets of natural language documents and code is turning our field on its head. The combination of clever architectures (e.g. transformers~\cite{GPT-3,https://doi.org/10.48550/arxiv.1706.03762,https://doi.org/10.48550/arxiv.2301.07069}) and huge amounts of data is creating a generation of pre-trained, relatively general-purpose models that seem to be able to make very accurate predictions of natural language as well as code. LLMs excel at tasks such as completing code and generating code from natural language descriptions ~\cite{10.1145/3520312.3534862}. One of the largest models, Codex ~\cite{chen2021evaluating}, has been used in the real-world tool Github Copilot~\cite{dakhel2022github} as an assistant for developers that can generate code based on user input. Github Copilot has the ability to produce code in various programming languages, given a specific context known as a prompt, which may include comments, method names, and existing code. Another one of those language models that offers a natural and intuitive interface for humans to interact with is ChatGPT,~\footnote{https://openai.com/blog/chatgpt/}
% Vanessa write a bit more about LLMs, prompt
which was made publicly available at the end of 2022. Since then, it has seen a meteoric rise in interest, some of which related to its capabilities in supporting software engineering tasks. 

\begin{figure}[]
\centering
\caption{Generated Commit Message from ChatGPT}
\label{fig:motivatingexample}
\resizebox{\columnwidth}{!}{%
\begin{tabular}{|l|}
\hline
\textbf{Diff (part of the prompt)} \\ \hline
\begin{tabular}[c]{@{}l@{}}--- a/core/.../CursorToBulkCursorAdaptor.java\\ +++ b/core/.../CursorToBulkCursorAdaptor.java\\ @@ -143,8 +143,7 @@ public final class\\ CursorToBulkCursorAdaptor ...\\ 
public void close() \{ \\ 
maybeUnregisterObserverProxy();\\ 
- mCursor.deactivate();\\ 
-\\ 
+ mCursor.close();\\ 
public int requery(IContentObserver observer, CursorWindow window) \{ 
\end{tabular} \\ \hline \hline
\textbf{ChatGPT's output} \\ \hline
\begin{tabular}[c]{@{}l@{}}~\textbf{Commit message:} Fix cursor closing method in CursorToBulkCursorAdaptor\\ ~\textbf{Justification: }The previous implementation was calling  the deactivate()\\ method on the cursor, which is no  longer the correct method for closing\\ a cursor in the current version of the Android framework. This change \\updates the code to use the close() method instead, ensuring that the cursor is \\ properly closed and resources are released.\end{tabular} \\ \hline \hline
\textbf{Original commit message} \\ \hline
\begin{tabular}[c]{@{}l@{}} Call close() instead of deactivate() in CursorToBulkCursorAdaptor.close()\\

This seems like a long-lived copy-and-paste bug.\\

Fixes http://b/issue?id=2015069\\

This is the master version of change \\
https://android-git.corp.google.com/g/8636 which never
made it into Eclair.
\end{tabular} \\ \hline
\end{tabular}%
}
\end{figure}

In a pilot test, we observed that ChatGPT seemed to be able to generate very high-quality commit messages that were descriptive of the changes when presented only with the diffs. Moreover, it also seemed to be able to generate justifications that seemed plausible for the changes. 
Figure~\ref{fig:motivatingexample} shows an example of a diff\footnote{https://github.com/allydev/android\_frameworks\_base/commit/\\d6dfca8302cc3ccc20113e90e7a65e4aad86fcac} casually entered into a ChatGPT conversation along with the prompt ``Generate a commit message for the following diff. Additionally, provide a justification for the changes.'' As seen, not only the short message is descriptive, but the justification seems surprisingly plausible, vaguely suggesting that it was due to an API change in the Android framework (``no longer the correct method ... in the current version ...''). ChatGPT's justification is surprising, because nothing in the input diff refers to the Android framework, much less to an API change. Moreover, the justification provided in the original human-written commit message is  different: a simple bug fix. ChatGPT seems to be drawing its prediction from a much broader, and hidden, context than that explicitly provided in the prompt. We posited that maybe the developer forgot to mention that there was an API change, not just a bug, and ChatGPT was able to infer the real reason behind the change.

To verify ChatGPT's justification, we looked at the project where the commit came from, and verified that, indeed, it is using the Android framework. We then looked at Android's Cursor class\footnote{https://developer.android.com/reference/android/database/Cursor}, and observed that the deactivate() method has, indeed, been deprecated. At this point, ChatGPT's justification was plausible.

But being plausible does not mean that it is true. We then fact-checked ChatGPT's justification by analyzing the history of the project where this commit was made and the Android framework, cross-checking the dates. It turns out that at the time of this commit (March 2010), the deprecated method was still very much part of Android's API (level 5, code name ECLAIR); deprecation only happened in July of 2012, several versions later (level 16, code name JELLY\_BEAN). ChatGPT's suggested commit message is high-quality and plausible, but ultimately inaccurate and misleading with respect to the real reason behind the change at the time, as explained by the original developer: a simple bug fix.

The combination of ChatGPT's very plausible commit messages and our discovery that at least some of them were misleading, or plain wrong, led us to design a more systematic empirical study. The overall objective of the paper is to explore the possibilities of ChatGPT in ACMG and to identify any challenges that need to be addressed to make it a viable solution.

This paper describes and reports the results of our systematic evaluation of ChatGPT for ACMG, comparing it to state-of-the-art research models that have been recently proposed in the literature. As baselines of comparison, we chose NNGen~\cite{9000000}, CoDiSum ~\cite{Xu2019CommitMG}, CommitBERT~\cite{jung2021commitbert}, and FIRA~\cite{9793882}.
% This level of detail doesn't belong here, it belongs later in the paper
%We picked these models because their authors made available the datasets, results, source code, and hyperparameters, which allows us to replicate the results of the original papers. We were unable to replicate CoRec~\cite{10.1145/3464689} or RACE~\cite{https://doi.org/10.48550/arxiv.2203.02700,exemplar} due to errors in the code and lack of response from the student authors. Additionally, attempts to replicate the Revisiting Learning-based Commit Message Generation~\cite{anonymous_2022_7042249} (presented at ICSE 2023) were unsuccessful as the student authors did not provide the preprint material. We also attempted to replicate the FIRA ~\cite{9793882} model, but did not have sufficient time.
The focus of our work is on qualitative evaluation of the \emph{correctness} of the outputs of all the models with respect to the actual changes, not with respect to the original commit message written by the developers. For ChatGPT, we provide a categorization of the mistakes we were able to identify, and discuss its implications for commit tasks.
We aim to answer the following research questions:
\begin{itemize}
\item \textbf{RQ1}: Is ChatGPT fit for purpose? 
\item \textbf{RQ2}: How does ChatGPT compare to the previous ACMG models? 
\item \textbf{RQ3}: What are the main classes of mistakes in ChatGPT's generated messages? 
\item \textbf{RQ4}: How do prompts affect the generated commit messages created by ChatGPT?
\end{itemize}
Our research makes the following contributions:
\begin{itemize}
    \item As far as we know, this is the first study on the correctness of commit messages with respect to the changes in the code. As powerful AI tools become more prevalent, this change of focus is critical.
    \item Comparatively, we show that zero-shot ChatGPT outperforms the baselines in all aspects. Moreover, we show that it is capable of generating the rationale behind the changes, which is something the baseline models cannot not do.
    \item We show that, with respect to the ultimate goal of commit tasks, ChatGPT is a viable model for generating commit messages and rationales that are as good as, and often better, than the human-generated ones. We also show that, when using it as zero-shot predictor, it sometimes makes mistakes, so it must not be blindly trusted.
    \item We provide an in-depth analysis and a simple categorization of ChatGPT's mistakes, and suggest prompt engineering mechanisms (\emph{aka} few-shot learning) for dealing with them.
\end{itemize}

The remainder of the paper is structured as follows.
In Section 2, we discuss related work on the different ACMG models proposed. In Section 3, we provide background information on large language models. In Section 4, we provide background information on ChatGPT. In Section 5, we outline our experiment design. Our experiment results are presented in Section 6. In Section 7, we discuss our findings.  Section 8 highlights the threats of validity to our study. Finally, in Section 9, we summarize our main findings and conclude the paper. 

\section{Related Work}
\noindent

This section provides an overview of the  different methodologies proposed in the literature for auto-generating commit messages.

$\textbf{Rule-based:}$ DeltaDoc~\cite{10.1145/1858996.1859005} employs templates such as "do X Instead of Z" to generate commit messages based on the control flow of the program between code versions. ChangeScribe~\cite{7203049,CortesCoy2014OnAG} proposes templates using method stereotypes~\cite{4021317} and commit stereotypes ~\cite{6080822}. Shen~\cite{7551998} used predefined templates to automatically generate commit messages. In general, the template-based have weak capability in describing the rationale and purpose of code changes. These templates are a specific format for the generated commit message, and can include specific metadata from the code revision. However, these approaches do not fully account for all situations or clearly convey the reasoning or intent behind code changes.

\noindent
$\textbf{Retrieval-based:}$ 
Retrieval-based methods leverage information retrieval techniques to adopt existing commit messages from code changes.  For example, given a code change as a query,  Liu~\cite{9000000} uses a nearest neighbor algorithm to select the most similar code change from the training set.  The training diff with the highest BLEU-4 score is regarded as the nearest neighbor of the new diff. Similarly, Huang ~\cite{10.1109/ESEM.2017.56,huang2020learning} uses both syntax and semantic similarity as the similarity metric. However, retrieval-based techniques are inaccurate when there are no  similar code changes in the retrieved database, as they can only output existing commit messages instead of generating new ones.  

\noindent
$~\textbf{Learning-based:}$ These methods have the capability to make good predictions for commit messages given source code diffs, and have shown promising results. They have become popular for solving the problem of generating commit messages.  Works such as Jiang et al.~\cite{8115626}, Loyola et al. ~\cite{loyola2017neural},  Liu et al.~\cite{Liu2019GeneratingCM} , Xu et al. ~\cite{Xu2019CommitMG}, and Pavel et al. ~\cite{Pavel2021ASN} have used neural machine translation to generate commit messages. Awad et al.~\cite{10.1145/3328833.3328873} used hidden markov models, while Jung ~\cite{jung2021commitbert} and Nie  ~\cite{nie2021coregen} use transformers. Despite their success, these methods tend to favor common words and produce less readable output, as well as suffering from exposure bias and and that they  generate high-frequency words but ignore low-frequency ones.

\noindent
$\textbf{Hybrid methods:}$ Liu et al.~\cite{liu2020atom} propose using an abstract syntax tree to represent code changes and integrating both retrieved and generated messages through a hybrid ranking system. This system prioritizes the most accurate message from both retrieved and generated messages for a specific code change. Wang et al.~\cite{10.1145/3464689} put forward a model that combines the advantages of retrieval- and learning-based methods for ACMG.  Their approach is capable of mitigating the exposure bias. Shia ~\cite{exemplar,https://doi.org/10.48550/arxiv.2203.02700}proposes a novel exemplar-based neural commit message generation model, which uses similar commit messages as examples to guide the neural network to generate informative and readable commit messages.  Dong et al.~\cite{9793882} proposed a new method for automatically generating commit messages for code changes. They represent the changes using detailed graphs and use machine learning to generate the messages. However, these methods still have limitations, particularly when compared to other methods such as those based on templates, retrieval, or learning. There is ongoing research in the area of combining different models to improve performance. Generating commit messages automatically is a complex task, as it involves converting structured code into natural language. There are many challenges involved, such as selecting the appropriate translations for words with multiple meanings, properly transliterating or translating named entities, and adapting the system to different domains and variations in language.

\section{Background on LLMs}

Large Language models (LLMs) trained on very large datasets have shown outstanding results in natural language processing tasks ~\cite{https://doi.org/10.48550/arxiv.1706.03762,PetroniRRLBWM19}.
As they become larger, they become more costly to train and update. This can make it difficult to adjust the model's behavior in real-world applications, particularly when it comes to factual knowledge ~\cite{geva2020transformer,https://doi.org/10.48550/arxiv.1909.01066,tang2022understanding} that plays a critical role in the model's performance. Petroni et al.~\cite{PetroniRRLBWM19} used LAMA probing to demonstrate that BERT models ~\cite{https://doi.org/10.48550/arxiv.2204.02311} function as knowledge bases by storing factual world knowledge. Roberts et al.~\cite{https://doi.org/10.48550/arxiv.2204.02311} establishes similar behavior for T5 models. However, since most factual knowledge is constantly changing, the stored information may be outdated or not include new facts, leading to incorrect or poor prediction ~\cite{lazaridou2021mind,https://doi.org/10.48550/arxiv.2205.02832}. The latest trend is to build on pre-trained LLMs and fine-tune them with specific tasks in mind. For example Codex~\cite{chen2021evaluating} builds on GPT-3~\cite{GPT-3} for the purposes of predicting source code given natural language prompts. A version of Codex powers GitHub's Copilot~\cite{dakhel2022github}.

\noindent
$\textbf{Mistakes of LLMs:}$
LLMs have been effective for various tasks, but are also known for producing hallucinated  content ~\cite{Ji_2022}. Hallucination refers to generating plausible statements that are factually incorrect.
They also suffer from temporal misalignment~\cite{mallen2022not}. Temporal misalignment refers to  when training and evaluation datasets are drawn from different time periods.
Sobania et al.~\cite{sobania2023analysis} conducted a study evaluating ChatGPT for bug fixing and found that it generated different responses based on the request. This is a result of the model's abstracted output, which can lead to hallucinations, a common issue with advanced NLP models. A study by Dzri et al.~\cite{dziri2022origin} showed that standard benchmarks contain over 60\% of false responses from these models, with some amplifying the hallucination. 
Additionally, LLMs have the potential to have bias representation. Chen et al.~\cite{chen2021evaluating} found that Codex can be prompted in ways that generate racist, derogatory, and otherwise harmful outputs as code comments and  found that code generation models raise further bias and representation issues beyond problematic natural language.

\noindent
$\textbf{Prompt Engineering:}$ Prompts provide a natural and intuitive interface for humans to interact with and utilize language model systems. It is a widely-used generic method for NLP tasks ~\cite{schick2020exploiting,GPT-3,sanh2021multitask}. Denny et al.~\cite{https://doi.org/10.48550/arxiv.2210.15157} conducted a study of prompting GitHub Copilot  to solve a series of programming questions. Copilot is able to  successfully solve around half of the problems on the first attempt. It is able to solve the remaining 60\% of the questions, with changes to the prompt description. Changing the form of the question that is asked of Copilot can influence the
correctness of the answer provided. Ross et al.~\cite{ross2023case} came up with a prototype of the Programmer's Assistant which integrates a chatbot with the code editor. The prompt needs to be engineered to get  specific class of result. LLMs need to be prompted carefully, either manually ~\cite{10.1145/3411763.3451760} or automatically ~\cite{shin2020autoprompt,zhou2022large}, as models don't comprehend the prompts in the same manner as humans do.

\section{ChatGPT}

ChatGPT is a chatbot developed by OpenAI that has been made publicly available on November 30, 2022. ChatGPT builds on OpenAI's LLM GPT-3~\cite{GPT-3}. GPT-3's main objective is to make accurate predictions of natural language, and ChatGPT is an impressive demonstration of those accurate predictions. However, ChatGPT \emph{seems} to include much more than GPT-3. Specifically, it \emph{seems} to have also been trained to predict software languages of many kinds (perhaps using Codex), as well as mathematical and scientific languages. \emph{Supposedly}, these additional trainings were made as fine-tunings of GPT-3. Recent \emph{news articles}\footnote{https://time.com/6247678/openai-chatgpt-kenya-workers/} also suggest that a large number of human knowledge workers have been used to generate training data.

The emphasized words of the previous sentences reflect a serious problem that we, as researchers, are facing: we do not know how exactly ChatGPT has been trained, and what techniques might have been used to make it seem like a qualitative leap from the previous generation of language models. Unfortunately, LLMs carry significant costs that no research institution has been, or will be, able to cover. The few papers published by OpenAI about GPT-3, Codex, etc. give some general indications of how it works, but the real power of ChatGPT \emph{seems} to come from the brute-force approach to solving problems: increase the number of machines for training and holding the weights, and increase the number of people for generating training data. For now, OpenAI offers access to their trained models' output through a human-facing Web interface and a machine-facing Web API; that is what we used in this work.

We do not know if ChatGPT's training data included datasets of diffs and commit messages, explicitly, and whether its base model was fine-tuned for the specific task of generating commit messages from diffs. It is clear, though, that ChatGPT is capable recognizing the syntax of diffs, explaining them back, and generating commit messages for them.

\section{Experiment Design}

\subsection{Baselines}~\label{ACMGmodels}

Our baselines are relatively small models trained specifically to predict commit messages from diffs, namely: NNGen~\cite{9000000}, CoDiSum ~\cite{Xu2019CommitMG},  CommitBERT~\cite{jung2021commitbert}, and FIRA~\cite{9793882}. These models were selected as baselines because they have available datasets and/or source code, which allows us to access or generate data for our qualitative analysis.\footnote{We were unable to use CoRec~\cite{10.1145/3464689} and RACE~\cite{https://doi.org/10.48550/arxiv.2203.02700,exemplar} due to errors in the code and lack of response from the authors. Additionally, attempts at using the Revisiting Learning-based Commit Message Generation (presented at ICSE 2023) were unsuccessful as the authors did not provide the preprint material.} Each of these models comes with their own datasets of input (some representation of diffs) and output (commit messages). In assessing the baselines, we used random samples of their own datasets.

A major difference between ChatGPT and the baselines of our study is that the baselines are relatively small models that can be trained by researchers in typical University computing infrastructures, while ChatGPT cannot. Another difference is that we know that these baselines have been specifically trained to generate commit messages from diffs, and we do not know if that happened for ChatGPT. Given these asymmetries of information, it is perhaps unfair to compare them directly. Nevertheless, it is important to study what ChatGPT can do with respect to commit messages, as well as its problems, and compare it to simpler models.

\subsection{Dataset}

\begin{figure}
    \centering
    \includegraphics[width=\columnwidth]{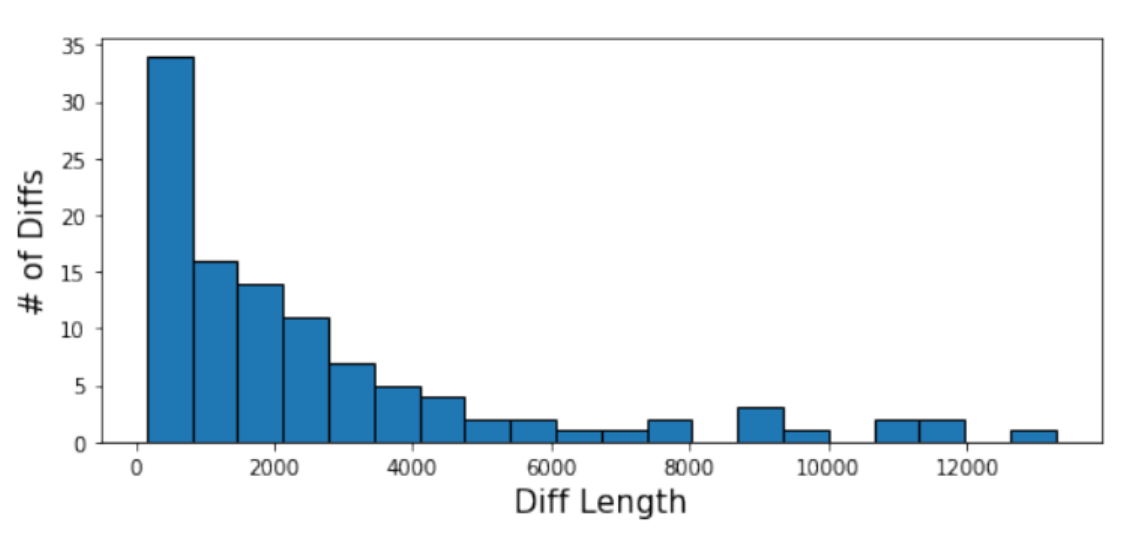}
    \caption{Distribution of diff size, measured in non-blank characters.}
    \label{fig:sizehist}
\end{figure}

To evaluate ChatGPT, we compiled a dataset of 108 diffs and their original commit messages. For a   very large population of code changes, 108 samples places our analysis on a confidence level of 95$\pm$9.4\%~\cite{hubert2010international}. The samples were randomly selected from two prior datasets, namely the dataset associated with Tian et al.'s work~\cite{tian2022makes} and from within the projects used by CommitBERT~\cite{jung2021commitbert}. Figure~\ref{fig:sizehist} shows the histogram of the size of the diffs, measured as the number of non-blank characters.

To obtain the generated commit messages from ChatGPT, we input the prompt ``Generate a commit message for the following diff and add a justification for the change:'' followed by the specific diff being evaluated. Every prompt with a diff is entered in a freshly created chat session, in order to avoid building up context from diff to diff. With this, we have 108 samples of diffs, their original commit messages, and the commit messages generated by ChatGPT.

\subsection{Methods}
The following describes the methods we used for answering the three research questions.

\subsubsection{RQ1: Is ChatGPT fit for purpose?}
\label{rq1}

One problem of these modern chat bots is that, given the exact same prompt, they do not generate the exact same reply. In order to check whether ChatGPT is capable of functioning as a reliable recommender of commit messages, we performed a competency test consisting of two parts: (1) an expertise test, with a line of questioning about commits, diffs, and examples of those; and (2) a consistency test where we gave it the exact same 10 diff prompts 10 different times using new chat sessions each time (so, no context); we then measured the similarity of its replies using BLEU, ROUGE-L, METEOR, and a BERT-based semantic similarity metric.\footnote{https://github.com/AndriyMulyar/semantic-text-similarity} We also performed a qualitative manual inspection.

\subsubsection{RQ2: How does ChatGPT compare to the previous ACMG models?}
\label{rq2}

The quantitative evaluation of the baseline models' accuracy is taken from their original papers. For ChatGPT, we calculate BLEU~\cite{10.3115/1073083.1073135}, ROUGE-L~\cite{Lin2004ROUGEAP}, and METEOR~\cite{banerjee-lavie-2005-meteor}.

For the qualitative evaluation, we take a random sample of 100 generated commit messages from all the baseline models combined (20 each, for a combined confidence level of 95$\pm$9.8\%), as well as 108 ChatGPT's messages, and manually analyse them in four dimensions:

\begin{itemize}
\item \textbf{Semantic similarity}: whether the generated messages convey the same general information as the original commit messages.
\item \textbf{Quality}: whether the generated messages contain a description of \emph{what} changed and \emph{why}~\cite{tian2022makes}.
\item \textbf{Correctness}: whether the generated messages describe facts inferred from the corresponding diffs.
\item \textbf{Improvement}: whether the generated messages are qualitative better than the original ones.
\end{itemize}

The classifications were made by some of the authors, together. We then performed an external validation of our judgements using the Delphi method~\cite{holey2007exploration}, which relies on using a pool of external experts to arrive at general conclusions. The criteria for expertise consisted of having used source control systems for at least 2 years. The Delphi method was implemented in the form of a survey deployed in a class of 41 computer science master students, of which 10 passed the inclusion criteria for expertise. The survey consisted of 5 questions each containing a randomly selected diff from the 200 samples in our study, along with the original message and the generated message. The experts were asked to judge whether the generated message was semantically similar to the original, whether it had quality, and whether it correctly reflected the diff. The survey included the definitions for these words. One of us, who was not involved in the qualitative evaluation, served as the moderator of this Delphi process. One round of the Delphi method was enough to converge, since for all 5$\times$3=15 judgements, only one of the authors' judgments (related to semantic similarity) differed from the judgement of the majority of external experts.

\subsubsection{RQ3: What are the main mistakes in ChatGPT's generated messages?}

We identify the mistakes ChatGPT makes in its generated messages by manually inspecting the 108 diffs and the original commit messages. 

%The result of this analysis is a set of categories that explains all mistakes that we were able to identify.

We analyzed the commit messages incrementally, that is, we qualitatively analyzed the commit message, following the open coding protocol~\cite{glaser2016open}. Two of us performed the analysis independently, coding whether there was mistake made by ChatGPT; if a mistake was identified, then we 
labeled it. This resulted in codes regarding types of mistakes ChatGPT made.  In several cases, we also analyzed the project of origin and its history. When doing the  analysis, each emerging code was compared with the existing codes to determine if the emerging code was a discrete category or a subset of an existing code. We carried out the whole procedure via continuous comparison throughout the coding sessions and through negotiated agreement~\cite{forman2007qualitative}. We discussed the rationale for applying specific codes, and reached a consensus during the negotiated agreement process. Two authors agreed on (1) whether it was a mistake, and (2) definition of the mistake.

%Since we did not know \emph{a priori} what we were looking for, we started with a pilot dataset of 80 commits and their messages. Some of these had incomplete diff information, some where complete, but most were very small. Using that pilot dataset, we had ChatGPT generate commit messages for them. We used the following methodology for arriving at the categories. One of the authors made a first pass at identifying mistakes and their types. A second made a second pass, and either agreed or disagreed with the classification. We calculated Inter-Rater Reliability (IRR), which was 95\%. After that, we had a tentative set of categories.

%However, rather than using that first dataset for the rest of the study, we discarded it, and created the study's main dataset of 108 commits referred to before, independent of the tentative set of categories we had created from the pilot. We then classified the mistakes in the main dataset according to the categories we had created before.

%In summary, a first dataset was used for our own training, and a second, larger, new dataset was used for testing the categories we learned from the first one.

\subsubsection{RQ4: How do prompts affect the generated commit messages created by ChatGPT?}

The following is an example of our baseline prompt:

\begin{lstlisting}
generate a commit message for the following diff, and add 
a justification for the change: 
diff --git a/pom.xml b/pom.xml
index 6fdbe7dc1cd..c62655cb229 100644
--- a/pom.xml
+++ b/pom.xml
@@ -57,7 +57,7 @@
     <okio.version>1.6.0</okio.version>
 
     <!-- Test Dependencies -->
-    <junit.version>4.11</junit.version>
+    <junit.version>4.12</junit.version>
   </properties>
 
   <scm>
\end{lstlisting}

We experimented with another zero-shot prompt that consisted of a long introductory text about the syntax of diffs, and compared the results with the baseline prompt. We also experimented with introducing omissions in the diffs, to check whether ChatGPT's messages would degrade.

Finally, we experimented with two-shot prompts: for the generated messages that were misleading or wrong, we followed up the conversation with second prompts that were carefully crafted for the type of mistake we identified, according to the categories of RQ2.

\section{Experiment Results}

\subsection{RQ1: ChatGPT's Fitness for Purpose}

The following is the chat log of the expertise test we performed on ChatGPT (CG): 

\begin{lstlisting}
Us>> Explain what a commit is in the context of source code version control.

CG>> A commit in the context of source code version control is a record of changes made to a set of files in a repository. Commits are usually made to a local repository, which is then pushed to a remote repository for sharing and collaboration. Each commit represents a new state of the codebase, and contains a message that describes the changes made. Commits serve as a way to keep track of the evolution of the codebase over time, and to allow developers to revert to previous versions if necessary. Commits also allow developers to collaborate on the codebase by merging their changes with those made by others.

Us>> And what is a diff?

CG>> A diff is a representation of the difference between two sets of code, typically two versions of the same file or two different files. A diff is usually shown in a human-readable format that highlights the lines that have been added, removed, or changed between the two versions of the code. The diff is used in version control systems to track changes to code over time and to help resolve conflicts that can arise when multiple people make changes to the same codebase.

Us>> Give me an example of a diff.

CG>> Here's an example of a diff:
--- a/file.txt
+++ b/file.txt
@@ -1,3 +1,5 @@
 This is the original content of file.txt.
-Here's a line that's been removed.
+Here's a line that's been added.
 And here's another line that's been added.
The --- a/file.txt and +++ b/file.txt indicate the original and modified versions of the file, respectively. The lines starting with @@ show the line range of the changes, with the - symbol indicating removed lines and the + symbol indicating added lines.

Us>> It seems you got the last line wrong. That line does not start with a `+`

CG>> I apologize, that was an error on my part. Here's a corrected example:
--- a/example.txt
+++ b/example.txt
@@ -1,3 +1,3 @@
 This is an example file.
-It has two lines.
+It has three lines.
 And this is the third line.

 Us>> Very good. What makes a good commit message? 

 CG>> A good commit message should be concise and descriptive, explaining the purpose of the changes made in the commit. It should include the following elements:
- A brief summary of the changes, using imperative tense (e.g. "Fix bug" rather than "Fixed a bug").
- Optionally, a more detailed explanation of the changes, providing context and reasoning for the changes made.
 - A reference to any relevant issues or tickets, if applicable.
Good commit messages are helpful for other contributors to understand the history of the project and make it easier to revert to a previous state if necessary. A clear and descriptive commit message also makes it easier to maintain a project over time.

Us>> Suggest a commit message for the following diff:
diff --git a/pom.xml b/pom.xml
index 6fdbe7dc1cd..c62655cb229 100644
--- a/pom.xml
+++ b/pom.xml
@@ -57,7 +57,7 @@
     <okio.version>1.6.0</okio.version>
 
     <!-- Test Dependencies -->
-    <junit.version>4.11</junit.version>
+    <junit.version>4.12</junit.version>
   </properties>
 
   <scm>

CG>> "Update JUnit version to 4.12 in pom.xml"
\end{lstlisting}

Although ChatGPT made one mistake, it corrected itself when the mistake was pointed out. The conversation made it clear that its training was enough for it to demonstrate knowledge about commits, diffs, and commit messages.

After this, we performed a consistency test of 10 diff prompts, each repeated 10 times, and measured the  similarity of ChatGPT's suggested commit messages for each prompt. Table~\ref{tab:consistency-test} shows the results. The numbers represent the similarity within the 10 instances of the 10 prompts. For example, prompt 1's 10 instances scored an average of 48 on BLEU, 32 on ROUGE-L, 30 on METEOR, and 64 on the BERT-based semantic similarity metric. This last metric is particularly meaningful, as it transforms words into BERT encodings that are then compared in that semantic space. These numbers show a very high syntactic and semantic similarity of ChatGPT's responses to each of the same prompts, a signal that the messages are not  completely inconsistent.

We further inspected the messages manually, and confirmed that, in general, the different responses to the same prompt convey the same general idea, even when the sentences and their ordering are different. For example, in 4 out of 10 cases, we observed that it picked a particular point to be the lead short description, and that choice was different; but, if it was not in the lead sentence, that point was present in the longer justification, so the overall semantics of the generated text was the same.

\begin{table}[]
\centering
\caption{Consistency Test (\%Similarity)}
\label{tab:consistency-test}
\resizebox{0.8\columnwidth}{!}{%
\begin{tabular}{|l|c|c|c|c|}
\hline
\textbf{Prompt} & \textbf{BLEU} & \textbf{ROUGE-L} & \textbf{METEOR} & \textbf{BERTSim}\\ \hline
1   &  48   &  32   & 30 & 64 \\ \hline
2   &  63   &  47   & 48 &  74 \\ \hline
3   &  55   &  49   & 43 &  74 \\ \hline
4   &  56   &  41   & 40 &  71 \\ \hline
5   &  51   &  37   & 34 &  68 \\ \hline
6   &  46   &  32   & 29 &  64 \\ \hline
7   &  55   &  41   & 39 &  74 \\ \hline
8   &  47   &  34   & 32 &  64 \\ \hline
9   &  46   &  33   & 39 &  67 \\ \hline
10  &  54   &  42   & 38 &  76 \\ \hline
\end{tabular}%
}
\end{table}

\noindent
\textbf{RQ1: Given these results, we conclude that ChatGPT is fit for the purpose of generating commit messages, because it demonstrates knowledge of the material, and its generated messages, although always different, demonstrate acceptable semantic consistency within its non-deterministic behavior.}

\subsection{RQ2: ChatGPT vs. Baselines}

\begin{table}[]
\centering
\caption{Quantitive Results (\%Similarity)}
\label{quantitive-results}
\resizebox{0.8\columnwidth}{!}{%
\begin{tabular}{|l|l|l|l|}
\hline
\textbf{Model} & \textbf{BLEU} & \textbf{ROUGE-L} & \textbf{METEOR} \\ \hline
NNGEN        & 16.42         & NR         & NR     \\ \hline
CoDiSum      &  16.55        &  9.73      & 12.83  \\ \hline
CommitBERT   & 13.32         & NR         & NR     \\ \hline
FIRA         & 17.67         & 21.58      & 14.93  \\ \hline
ChatGPT      &  22.8       & 20.8      & 15.19   \\ \hline
\end{tabular}%
}
\end{table}

The quantitative results are shown in Table~\ref{quantitive-results}. The numbers represent the similarity between each model's generated messages and the original commit messages in their own datasets. For ChatGPT, we used our dataset of 108 diffs/messages. 

According to these results, ChatGPT performs better than the baselines on the BLEU score, slightly better on the METEOR score, and worse than FIRA on the ROUGE-L score. However, all five models are unremarkable, as the scores are relatively low. We will come back to discussing the importance of these metrics after presenting the qualitative results.

\begin{table}[]
\centering
\caption{Qualitative Results (\%)}
\label{qualitative-results}
\resizebox{0.8\columnwidth}{!}{%
\begin{tabular}{|l|c|c|c|c||c|} \hline
  & \textbf{Sim} & \textbf{Qual} & \textbf{Corr} & \textbf{Improv}& \textbf{Med}\\ \hline
Baselines & 21 &  0   & 13    &   0 & 19 \\ \hline
ChatGPT   & 66  & 100 &  83 & 42& 381 \\ \hline
\end{tabular}%
}
\end{table}

Table~\ref{qualitative-results} presents the results of our manual qualitative evaluation of 208 commits and corresponding generated messages, 108 from ChatGPT and 100 from the four baseline models. The columns on this table are: semantic similarity (Sim), quality (Qual), correctness (Corr), and improvement (Improv), as defined in the previous section. The final column, Med, is the median of the length of the generated messages, measured in non-whitespace characters.

The qualitative differences between ChatGPT and the baselines are striking.
The generated commit messages from the baseline models are very short (median of 19 characters), and either vague or fail to capture the code changes. Only 21 out of 100 were semantically similar to the original messages. None of them provided a justification for the change (quality) beyond implicit action keywords such as ``fix,'' which was often incorrect. Worse, only 13 out of 100 were correct, in the sense that they were an accurate summarization of the changes in the diff; 87 were just wrong. Also, none of them was an improvement over the original commit message. Table~\ref{tab:sample-messages-baselines} shows eight examples of original and generated commit messages of the baseline models.

\begin{table}[]
\centering
\caption{Samples from the baseline models}
\label{tab:sample-messages-baselines}
\resizebox{\columnwidth}{!}{%
\begin{tabular}{|l|l|}
\hline
  \textbf{Original}  & \textbf{Generated} \\ \hline
Fix TsExtractor tests & Adds lintOptions to Library \\ \hline
Added joscar JAR &  Added missing JNA jar  \\ \hline
Recycle cached views when clearing the cache. & fix bug in store cached \\ \hline
Bring documentation of Service & fix typo \\ \hline
Change callback defaults for transfers & Fixes typo \\ \hline
Updated version number to match website & Fixes typo \\ \hline
add additional loggin & improve assertion message \\ \hline
fix the title & fix a typo \\ \hline
\end{tabular}%
}
\end{table}

In contrast, ChatGPT generates complete commit messages and explanations for code changes, as requested by our prompt. The text is grammatically correct, the commit messages are always in good form, and the justifications are always present (quality), resulting in suggestions with a median size of 381 characters. 66\% are semantically similar to the original commit messages and 83\% of them are correct. Even more surprising is that almost half of them (42\%) are better than the original commit messages, and many  could be considered better if we overcome some of the problems identified in the next section. Table~\ref{tab:sample-messages-ChatGPT} shows a couple of examples of excellent commit messages generated by ChatGPT.

\begin{table}[]
\centering
\caption{Samples from ChatGPT}
\label{tab:sample-messages-ChatGPT}
\resizebox{\columnwidth}{!}{%
\footnotesize
\begin{tabular}{|p{0.8\columnwidth}|}
\hline
 \textbf{Original}   \\ 
 Internal change \\\hline
 \textbf{Generated} \\
Fix typo in MaskRCNNLosses and improve box loss computation

The commit fixes a typo in the variable name "groundturth" in the function signature of call, and changed it to "groundtruth"
The commit also improves the computation of the box loss in the \_rpn\_box\_loss method by only considering valid boxes (boxes with at least one coordinate non-zero) in the loss calculation, instead of all boxes. The valid mask is created using the reshaped box\_targets, and is used to weight the Huber loss. The loss is then normalized by the sum of non-zero weights and additional normalizer provided by the function caller. This should result in a more accurate and stable box loss calculation.
\\ \hline \hline
 \textbf{Original}   \\ 
 Replaced sort with sorted. `sort` throws an error, so replaced with `sorted` which solves the same purpose \\ \hline
 \textbf{Generated} \\
 Fix sorting method in various code blocks

This diff changes the sorting method used in various code blocks from the "sort" method to the "sorted" method. This is likely done to ensure that the data is always sorted in a consistent manner, as the "sort" method sorts the data in place, while the "sorted" method returns a new sorted list. Additionally, the "sorted" method is a built-in function, which is likely more efficient than a custom sort function.
\\ \hline
\end{tabular}%
}
\end{table}

\noindent
\textbf{RQ2: With respect to the baselines, ChatGPT increases the number of correct commit messages 6-fold, and is capable of generating messages that are better than the original ones in 42\% of time, something that the baselines cannot do.}

After making this large qualitative evaluation of 208 commit messages, we make two final observations:

\noindent
\textbf{Poor quality messages.} The majority of human-written commit messages we analyzed were short -- an observation that has been measured before~\cite{10.5555/2486788.2486844}. Since we looked at the diffs, we found many situations where the commit messages written by developers were incomplete, imprecise, inaccurate, hard to understand, or simply wrong. That is the reason why so many of ChatGPT's generated messages (42\%) were better than the originals -- the human-written messages had many problems. That is also one of the motivations behind automatic commit generation research.

\noindent
\textbf{Metrics.} Referring back to the quantitative results shown in Table~\ref{quantitive-results}, we stated that they were unremarkable across the board. They do not correlate with our qualitative analysis, which shows a remarkable qualitative leap of ChatGPT with respect to the baselines. These metrics, which are widely used for measuring the performance of NLP models, do not measure semantic similarity, but simply a very narrow concept of similarity based, for the most part, on equality of words (METEOR includes synonyms, but that is as far as it goes). LLMs like ChatGPT seem to be quite creative with words, and therefore it is unlikely that they predict the exact same words as those chosen by developers on specific situations. These metrics are inappropriate, and 
this has been pointed out before~\cite{Dey2022EvaluatingCM,ma2022multi}.

These two observations, together, lead us to make a stronger statement about research in ACMG: we cannot rely on non-curated human-written messages as the ground truth. There are cases where the semantic similarity between ChatGPT's message and the original message is very low, but where ChatGPT is correct and the original message is incorrect, incomplete, or meaningless (see Table~\ref{tab:sample-messages-ChatGPT}). Work that uses NLP metrics in a way that optimizes for mimicking the original messages will be misguided by those metrics.

\subsection{RQ3: Categories of Mistakes}

In spite of its surprisingly good performance, ChatGPT also makes mistakes. As explained before, we followed an open coding protocol to come to a consensus about categories of mistakes. The following are the categories of mistakes that emerged from our open coding protocol: 
\begin{itemize}  
\item \textbf{Lack of context}: The message is missing the point, because it lacks project-specific context. For example, links to bug reports or some piece of historical information (e.g. the commit coming after a code review). 
\item \textbf{Truism}: The message (typically the justification) includes statements that are trivially true. For example, ``The upgrade ensures that the project has access to the latest bug fixes and features of the library, and helps to keep the dependencies of the project up-to-date.''
\item \textbf{Perceptual error}: The mistake seems to come from a misunderstanding of the diffs. For example, when the diff is too long and varied, ChatGPT tends to look only at some parts, usually the beginning, and may miss the main change. But it can happen also with very small diffs of just one line change.
\item \textbf{Hallucination}: The message (typically, the justification) makes too many unsubstantiated statements. For example, it justifies the change as a security issue, when it had nothing to do with security. The motivating example of the introduction is another instance of this mistake.
\item \textbf{Literal interpretation}: Failure to distinguish code from text, interpreting the words of the text as the changes themselves. This happened for some text (non-code) files: their changed lines described something, for example, a piece of code in documentation, that ChatGPT interpreted as being a change in the code.
\end{itemize}

Table~\ref{tab:mistakes} shows the number of ChatGPT's generated messages that had these different kinds of problems. Some messages had more than one problem. We noticed that there is a small correlation between lack of context and truism mistakes: when ChatGPT is missing context, it tends to ``fill up the space'' of the justification with truisms.

\begin{table}[]
\centering
\caption{Classification of ChatGPT's Mistakes}
\label{tab:mistakes}
\resizebox{0.6\columnwidth}{!}{%
\begin{tabular}{|l|r|} \hline
\textbf{Type of mistake} & \textbf{\# (\%)}    \\ \hline
Lack of context        & 38 (35\%) \\ \hline
Truism                 & 12 (11\%) \\ \hline
Perceptual error       & 10 (9\%)  \\ \hline
Hallucination          & 6 (6\%)   \\ \hline
Literal interpretation & 3 (3\%)   \\ \hline
\end{tabular}%
}
\end{table}

%...correlation between mistakes and size of diff...
\noindent
\textbf{RQ3: We identified five major categories of mistakes that seem to explain all of the problems we observed in the generated messages.}

\subsection{RQ4: Effect of Prompts}

In general, the prompts have a significant effect on ChatGPT's responses. We studied this effect along two dimensions: (1) incomplete zero-shot prompts; (2) two-shot prompts for response improvement.

\subsubsection{Incomplete zero-shot prompts}

The goal of this part of the study was to analyze how ChatGPT reacted to diffs that were missing seemingly useless information that has been eliminated from the input of some of the baselines. Specifically, NNGen's input omits the line range of the changes (e.g. ``@@ -1,3 +1,5 @@''), and CommitBERT and FIRA use only the added and deleted tokens, throwing away everything else.

We took 42 randomly selected diffs from NNGen's dataset, and fed them to ChatGPT under our baseline prompt. These diffs were just missing the line ranges, everything else was there. The performance degradation was very visible: in these 42 diffs, the semantic similarity between the generated messages and the original is comparable to the that of the prompts with complete diffs (67\%), but the correctness of the generated  messages is only 45\% (in contrast to 83\% on our dataset). Most of the errors are perceptual (26\%, in contrast with 9\% on our dataset), followed by truisms (24\%, in contrast to 11\%), followed by hallucinations (21\%, in contrast to 6\%), followed by lack of context (14\%, in contrast to 35\%), and finally literal interpretations (7\%, in contrast to 3\%).

Similarly, we made a few ad-hoc experiments with inputs similar to those used in CommitBERT and FIRA, with only additions and deletions, and nothing else, but it was clear that ChatGPT's predictions on them were mostly erroneous.

These experiments lead us to the conclusion that ChatGPT did, indeed, ``learn'' about diffs and is capable of correctly identifying them and of making good predictions of commit messages. But when the supposed diffs do not conform to the form that it ``learned,'' its performance degrades rapidly.

\subsubsection{Two-shot prompts}

This part of the study is a feasibility study for an interactive tool that automatically recommends commit messages, but where the problematic recommendations could be corrected by developers. We envision the creation of these messages as an interactive conversation. With that goal in mind, we wanted to find out if, and how, ChatGPT was capable of automatically correcting the problematic commit messages with simple factual feedback from the developer committing the changes. 

We selected diffs from each mistake category, for a total of 22 diffs, and we repeated the baseline prompts for them. As expected, ChatGPT gave slightly different responses, but with the same general ideas and the same mistakes. We then prompted it a second time with additional information that might correct the problems. Depending on the type of mistake, that information either pointed out a fact or added context data. Table~\ref{tab:prompt2} shows prototypical examples of these second prompts. The following is a concrete example of one of these interactions:

\begin{lstlisting}
US>> Generate a commit message for the following diff and provide a justification for the changes:
<the diff>

CG>> Commit Message: Refactor DefaultFuture to improve log message for timeout
Justification: The change refactors the DefaultFuture class to improve the log message for timeout events by removing sensitive data from the request object when logging. This change helps to maintain the privacy and security of user data by removing it from the log messages. The function getRequestWithoutData has been added to return a copy of the request object with its data field set to null.

Us>> Also, issue #5149, reduce log size. Try again.

CG>> Commit message: Fix #5149, reduce log size in DefaultFuture
Justification: This change resolves issue #5149 by reducing the size of logs generated by DefaultFuture. The change replaces the request object in the timeout message with a new request object where the data property is set to null, preventing sensitive data from being logged. A new method, getRequestWithoutData, has been added to return a new request object with data set to null.
\end{lstlisting}

The results are shown in Table~\ref{tab:two-shot-prompts-results}. As shown, after the second prompts, ChatGPT was capable of producing better commit messages that fixed all the problems, but sometimes it introduced new problems that had not been there before. 

\begin{table}[]
\centering
\caption{Examples of Second Prompts}
\label{tab:prompt2}
\resizebox{\columnwidth}{!}{%
\begin{tabular}{|l|p{0.8\columnwidth}|}
\hline
\textbf{Type of mistake} & \textbf{2$^{nd}$ Prompt Example}  \\ \hline
Lack of context        & \emph{Also, align with latest Spring Data Kay snapshots. And issue \#gh-7461. Try again.}  \\ \hline
Truism                 & \emph{Make it shorter. Try again.}  \\ \hline
Perceptual error       & \emph{volatile was also removed. Try again.} \\ \hline
Hallucination          & \emph{Also, issue \#5149, reduce log size. Try again.}  \\ \hline
Literal interpretation & \emph{This was just a text file. Try again.} \\ \hline

\end{tabular}%
}
\end{table}

\begin{table}[]
\centering
\caption{Results of ChatGPT's Mistake Correction }
\label{tab:two-shot-prompts-results}
\resizebox{\columnwidth}{!}{%
\begin{tabular}{|l|c|c|c|}
\hline
\textbf{Type of mistake (count)} & \textbf{Fixed} & \textbf{Fixed but new problems} & \textbf{Not fixed} \\ \hline
Lack of context  (5)        &  5   &  0  & 0 \\ \hline
Truism  (5)                 &  5   &  0  & 0 \\ \hline
Perceptual error (5)        &  2   &  3  & 0 \\ \hline
Hallucination (4)           &  2   &  2  & 0 \\ \hline
Literal interpretation (3)  &  0   &  3  & 0 \\ \hline

\end{tabular}%
}
\end{table}

Before we settled on our second prompts shown in Table~\ref{tab:prompt2}, we experimented with many approaches for error correction, most of which failed to produce the desired results. Specifically, prompts that tried to say too much failed to lead ChatGPT towards better messages; they seemed to add confusion to its prediction process. The most successful prompts were the very short ones that pointed exactly at what needed to be corrected, added, or removed.

The baseline prompts are always the same, and can be issued automatically upon commit actions. The second prompts, however, require conversation with the developer to gather context and feedback. Given how sensitive ChatGPT is to the exact formulation of prompts, it is not clear that a free-form conversational tool would be successful in practice; developers may prefer to edit the suggested message themselves, and fix them. But this is an empirical question about future tools that falls out of scope of this paper. This part of our study shows that it is possible to lead ChatGPT to correct many of its mistakes, but this requires experience and mastery of the English language.

\noindent
\textbf{RQ4: The prompts have a significant effect on ChatGPT's responses. Small omissions in the diffs lead to a noticeable increase in the number of mistakes. Additionally, in two-shot scenarios designed to produce better commit messages, the exact wording of the second prompt has a significant effect on whether a better message is generated or whether it gets worse. We found that very short commands that address the mistakes directly work better than more verbose prompts that seem to confuse ChatGPT's predictions.}

\section{Discussion}

We want to come back to the point about metrics. There have been several studies that quantify commits in large datasets~\cite{9609189,tao2022large}, and one study that goes qualitatively deeper~\cite{tian2022makes} with respect to the \emph{form} of commit messages. As far as we know, our work is the first study focusing on the \emph{correctness} of commit messages with respect to the respective diffs. We manually inspected almost 108 commits of various sizes and verified whether both the original commit messages and the generated ones truly captured the changes, many times following through to the original projects to better understand the context. This was a time-consuming process, but the findings were enlightening. Independent on our findings about ChatGPT, it was clear that the original commit messages had many problems; except for the trivial changes, it was rare to find a commit message that accurately summarized the nature of the change. We found many that were truly bad. ACMG work, so far, has focused on measuring and reporting NLP metrics that simply assume that the original commit messages are the gold standard to strive for. Based on what we saw, we believe this is a misguided goal. ACMG should strive to generate commit messages that are much better than those written by developers. 

With that more ambitious goal in mind, LLMs such as ChatGPT seem to be the answer to getting there. LLMs are not trained on [bad] examples of commits. Instead, they are trained to predict natural and artificial languages really well. In doing that, they seem to acquire reasoning capabilities that help in solving a large variety of knowledge-based tasks, such as ACMG.

Because generating formally good summaries from diffs is now a reality, the challenge from here on is to detect when the summary is incorrect or incomplete. This ties back with metrics, again. We may have been able to work around poorly-written commit messages from human developers, because, as last resort, we can ask them directly what they meant. But we cannot accept that from automated tools.

\section{Threats to Validity}

The main threat to the validity and repeatability of our findings is the fact that ChatGPT is  outside of our control, and cannot be externally kept for archival purposes. For example, on January 30, 2023, during the writing of this paper, OpenAI announced that a new version of ChatGPT had been deployed with ``improved factuality and mathematical capabilities.''\footnote{https://help.openai.com/en/articles/6825453-chatgpt-release-notes} To mitigate this, we redid some of our prompts to verify whether some of the problems we identified had been resolved with this update. It appears they have not.

In spite of the high IRR in our classification of ChatGPT's mistakes, our judgements can be questioned. We mitigate this by making our data available along with the paper. We believe the classifications are non-controversial in the vast majority of cases. 

Another threat to validity is the possibility of mistakes in running the scripts to replicate the four ACMG models. To mitigate this threat, we had multiple people check and verify that there were no errors.

%Our survey had a limited number of questions and was based on a limited sample of code changes selected from each project. This was due to the resource-intensive nature of evaluating each code change and its associated commit message and generated commit message. In our manual evaluation of the generated commit messages from the four baseline models, we only looked at 80 code changes and their corresponding commit messages and generated commit messages, as the process was time-consuming. The study was further restricted by a rate limit that only permitted feeding 80 code changes into ChatGPT at a time. Another limitation, was we did not know where the code diffs came from as, the dataset did not give the project.Another limitation was that we were unable to determine the source of the code differences, as the dataset for NMT, NNGen, CoDiSum, and FIRA  did not provide information about the project.

\section{Conclusion}

To help programmers write comprehensive commit messages, many automatic message commit models have been proposed. However, these models have limitations. Rule-based models do not cover all scenarios or effectively explain the reasoning behind code changes. Retrieval-based models may be inaccurate if no similar code changes exist in the database, only repeating existing messages instead of generating new ones. Learning-based models often prioritize common words, resulting in less readable messages, and also suffer from exposure bias by generating frequent words and neglecting less frequent ones. ChatGPT, a newly introduced large language model, can generate a recommended commit message by providing a prompt and the related code change.

We show that ChatGPT has much better performance than the ACMG models that have been proposed before. However, some of the suggested commit messages have problems, such as lack of context and unsubstantiated statements. Experiments using ChatGPT's dialogue show that it is possible to correct ChatGPT's mistakes using very simple and direct commands. 

Despite its great performance, the question remains whether the mental effort required to verify ChatGPT's answers outweighs its advantages. Incorporating automated approaches to provide ChatGPT with hints and verifying its responses, for example, by connecting it to the Github repository, could be a useful tool for software developers in their daily tasks. We hope our results and impressions will be helpful for future work with ChatGPT.

LLMs are redefining research in software engineering. Just a few months ago, ACMG looked both an important problem that could benefit from automation, and a hard problem to solve. The machine learning models proposed very recently in the literature were showing a steady, but incremental, progress towards the goal. ChatGPT clearly outperforms them in all dimensions. The reasons for why this happens are worth pondering. Brute force works, and that is something academic researchers cannot do. The question now is not how to develop methods for ACMG, but how to leverage and control LLMs towards that specific tasks.

\section{Data Availability}

All data collected for this study, as well as our statistical analyses, are made available as supplemental material. We intend to make the data and our analysis scripts publicly available upon publication of the paper.

\bibliographystyle{ACM-Reference-Format}
\bibliography{main}
\end{document}